\documentclass[prc,aps,showpacs,twocolumn]{revtex4}
\usepackage{amsmath}
\usepackage{amssymb}
\usepackage{graphicx}
\usepackage{color}
\makeatletter

\newcommand{\bs}[1]{\ensuremath{\boldsymbol{#1}}}
\renewcommand{\rm}[1]{\mathrm{#1}}
\newcommand{\bra}[1]{\big< \,{#1}\, \big| }
\newcommand{\ket}[1]{\big| \,{#1}\, \big> }
\newcommand{\braket}[2]{\big< \,{#1}\, \big| \,{#2}\, \big> }

\newcommand{\expect}[1]{\big< \, {#1} \, \big>}

\newcommand{\matrixe}[3]{\big< \,{#1}\, \big| \,{#2}\, \big| \,{#3}\, \big> }

\newcommand{\vek}[1]{\!\vec{\,#1}}

\newcommand{\half}{{\textstyle \frac{1}{2}}}

\renewcommand{\c}{\mathbf{c}}
\newcommand{\x}{\mathbf{x}}
\newcommand{\HF}{\mathrm{HF}}
\newcommand{\MF}{\mathrm{MF}}
\newcommand{\DF}{\mathrm{DF}}
\newcommand{\E}{{\cal E}}
\newcommand{\R}{\widehat{\cal R}}
\newcommand{\HM}{\widehat{\cal H}_{\mathrm{MF}}}

\newcommand{\pp}[1]{\frac{\partial}{\partial {#1}}}
\newcommand{\pd}[2]{\frac{\partial{#1}}{\partial {#2}}}
\newcommand{\dd}[2]{\frac{d{#1}}{d{#2}}}
\newcommand{\ddt}{\frac{d}{dt}}
\makeatother

\begin{document}
\title{On $^{229}$Th and time-dependent fundamental constants}

\author{Elena Litvinova$^{1,2}$, Hans Feldmeier$^{1}$}
\affiliation{$^1$GSI Helmholtzzentrum f\"ur Schwerionenforschung, Planckstr.~1,
             64291 Darmstadt, Germany\\
             $^2$Institute of Physics and Power Engineering, 249033 Obninsk, Russia
}
\author{Jacek Dobaczewski}
\affiliation{Institute of Theoretical Physics,
University of Warsaw, Ho\.za 69, PL-00-681 Warsaw, Poland}
\affiliation{Department of Physics, P.O. Box 35 (YFL),
FI-40014 University of Jyv\"askyl\"a, Finland}
\author{Victor Flambaum}
\affiliation{School of Physics, University of New South Wales, Sydney, NSW 2052, Australia}
\date{\today{}}

\begin{abstract}
The electromagnetic transition between the almost degenerate 5/2$^+$ and $3/2^+$ states in
$^{229}$Th is deemed to be very sensitive to potential changes in the fine structure
constant $\alpha$.  State of the art
Hartree-Fock and Hartree-Fock-Bogoliubov calculations are performed to
compute the difference in Coulomb energies of the two states which determines
the amplification of variations in $\alpha$ into variations of the transition frequency.
The kinetic energies are also calculated which reflect a possible variation in the nucleon
or quark masses.
A generalized Hellmann--Feynman theorem is proved including the use of density-matrix
functionals.
As the two states differ mainly in the orbit occupied by the last unpaired neutron
the Coulomb energy difference results from a change in the nuclear polarization
of the proton distribution. This effect turns out to be rather small and to depend
on the nuclear model, the amplification varies between about
$-4 \times 10^4$ and $+4 \times 10^4$.
Therefore much more effort must be put into the improvement of the nuclear models
before one can draw conclusions from a measured drift in the transition frequency
on a temporal drift of fundamental constants.
All calculations published so far do not reach the necessary fidelity.
\end{abstract}

\maketitle
\section{Introduction}

Measurements of a suspected temporal variation of the fine structure constant
by means of atomic transitions have reached an limit for $\delta \alpha/\alpha$ of less than
$10^{-16} \rm{yr}^{-1}$ \cite{RHSC08,FABD07,PLSS04,FKZH04,PrTM95,Flam08}.

A tempting idea to increase the sensitivity limit is to use the transition between
two states with very different Coulomb energies, because the drift in transition
frequency
\begin{equation} \label{eq:amplificationfactor}
\frac{\delta\omega}{\omega}=A \; \frac{\delta\alpha}{\alpha}
\ \ \ \rm{with}\ \ \ A= \frac{\Delta V_C}{\omega}
\end{equation}
is given by an amplification factor $A$ times the drift in $\alpha$.
Using the Hellmann--Feynman theorem $A$ is the ratio of the difference $\Delta V_C$ in
Coulomb energies of the two states involved in the transition divided by
the transition frequency $\omega$.
A promising candidate for that is the transition between the $3/2^+$ isomeric
state of the nucleus $^{229}$Th to its $5/2^+$ ground state \cite{PeiT03,Flam06}.
Recent measurements
yield $\omega=7.6\pm 0.5$~eV \cite{BBBB07} which on nuclear energy scales is an
accidental almost degeneracy.
Typical Coulomb energies for this nucleus are of the order
of $V_C \approx 10^9$~eV so that even a small difference $\Delta V_C$ could result
in a large amplification.

A big drawback of this idea is that the Coulomb energies cannot be measured
but have to be calculated with sufficient accuracy in a nuclear model.
Differences in charge radii and quadrupole moments of the nuclear states
are in principle experimentally accessible. Their measurement would reduce the uncertainty.

In a simplified picture the two states differ by the occupation of the last
neutron orbit.
The change in the Coulomb energy is thus due to a modified neutron distribution
in the excited state
that will polarize via the strong interaction the proton distribution
in a slightly different way than in the ground state.

In this paper we investigate in how far state of the art nuclear models
can provide reliable answers. For that in section \ref{sec:HellmannFeynman}
we revisit first the Hellmann--Feynman theorem
proposed by H.~Hellmann \cite{Hell33} and also by
R.~Feynman \cite{Feyn39} and show that it is also valid in a wider context
including density-matrix functional theory employed in nuclear physics.
After introducing the nuclear models in Sec.~\ref{sec:models}
we discuss in Sec.~\ref{sec:amplification} the amplification of the temporal drift of
fundamental constants like the fine structure constant or the nucleon mass
when monitoring transition frequencies.

For the nuclear candidate $^{229}$Th we perform in Sec. \ref{sec:results}
self-consistent nuclear structure calculations and determine via the
Hellmann--Feynman theorem the derivatives of the transition frequency
with respect to the fine structure constant and the nucleon mass.
A critical assessment is made on the predictive power when
calculating observables other than the energy.
Finally we summarize.

\section{Hellmann--Feynman theorem revisited}\label{sec:HellmannFeynman}
In this section we show that the Hellmann--Feynman theorem
holds not only for eigenstates of the Hamiltonian but also for
all stationary solutions in approximate schemes, provided they
are variational. This is of importance as the models we are using are
based on density-matrix functionals.

Let the Hamiltonian $H(\mathbf{c})$ depend on a set of
external parameters $\c=\{c_1,c_2,\dots\}$,
like the strength of an interaction or the mass of the particles.
The steady state solutions of the time-dependent Schr\"odinger equation
are given by the eigenvalue problem
\begin{equation}\label{eq:Schroedinger}
H(\mathbf{c})\ket{\Psi_n;\mathbf{c}}=E_n(\c)\ket{\Psi_n;\mathbf{c}}\ .
\end{equation}
Both, the energies $E_n(\mathbf{c})$ and the eigenstates $\ket{\Psi_n;\mathbf{c}}$
depend on $\mathbf{c}$.
When discussing intermolecular forces in 1933 H.~Hellmann \cite{Hell33} and
later in 1939 R.~P.~Feynman \cite{Feyn39} showed that a small variation
of an external parameter in the Hamiltonian (distance between nuclei)
leads to a change in the energy given by:
\begin{equation}\label{eq:HF}
\pd{}{c_i} E_n(\c)=\frac{\matrixe{\Psi_n;\c}{\pd{}{c_i}H(\c)}{\Psi_n;\c}}
                               {\braket{\Psi_n;\c}{\Psi_n;\c}} \ .
\end{equation}
The proof hinges on $\ket{\Psi_n;\c}$ being
an eigenstate of the hermitian $H(\c)$:
\begin{align}\label{eq:proof1}
\pd{}{c_i}& E_n(\c)=\pd{}{c_i}\frac{\matrixe{\Psi_n;\c}{H(\c)}{\Psi_n;\c}}
                               {\braket{\Psi_n;\c}{\Psi_n;\c}}\nonumber\\
       &\!\!=\frac{\matrixe{\Psi_n;\c}{\pd{}{c_i}H(\c)}{\Psi_n;\c}}
                               {\braket{\Psi_n;\c}{\Psi_n;\c}}\\
    &\!\!+\left[\frac{\matrixe{\pd{\Psi_n}{c_i};\c}{H(\c)}{\Psi_n;\c}}
                {\braket{\Psi_n;\c}{\Psi_n;\c}}\ +\ h.a.\right]\nonumber\\
  &\!\!-\left[\frac{\matrixe{\Psi_n;\c}{H(\c)}{\Psi_n;\c}}
                               {\braket{\Psi_n;\c}{\Psi_n;\c}}
  \frac{\braket{\pd{\Psi_n}{c_i};\c}{\Psi_n;\c}}
                               {\braket{\Psi_n;\c}{\Psi_n;\c}}\ +\ h.a.\right].
                               \nonumber
\end{align}
Inserting \eqref{eq:Schroedinger} in the terms enclosed in square brackets
leads to a cancellation of these terms and we are left with
Eq. \eqref{eq:HF}, the Hellmann--Feynman theorem.
The partial derivative of any eigenvalue $E_n(\mathbf{c})$ with respect
to a parameter $c_i$ is equal to the expectation value of the
partial derivative of the Hamiltonian calculated with the corresponding eigenstate
$\ket{\Psi_n;\mathbf{c}}$.

We now show that the same statement holds for any extremal point in an energy
functional. Let ${\cal E}(\c,\x)$ be the energy of a physical
system that depends on external parameters $\c$ and on a set of variational
parameters $\x=\{x_1,x_2,\cdots\}$ which characterize the state
of the system. $\x$ may also represent a set of functions in which
case partial derivatives are replaced by functional derivatives.

For example, in the Hartree-Fock approximation $\x$ would be
the set of occupied single-particle states that form a Slater determinant.
In an energy density functional $\x$ could be the local density $\rho(\vek{r})$,
and so on.

Steady state solutions $\x^{(n)}(\c),\ n=0,1,2,\dots$ of the system are obtained by the condition
\begin{equation}\label{eq:stationary}
0=\pd{\E}{x_k}(\c,\x)
\end{equation}
At the stationary points the energy assumes the values
\begin{equation}
E_n(\c)= \E(\c,\x^{(n)}(\c)),\ \ \ n=0,1,2,\dots  \ .
\end{equation}
Both, the energies and the parameters $\x^{(n)}(\c)$ characterizing the
stationary states depend on the constants $\c$.
In the ground state given by $\x^{(0)}$ the energy $\E(\c,\x)$
is in an absolute minimum with respect to variations in $\x$, while
the other possible solutions $\x^{(n)}, n\neq 0$, represent saddle points.

A variation of the external parameters at the stationary points leads to
\begin{align}\label{eq:variation}
\delta E_n(\c) &= \sum_i \pd{}{c_i}E_n(\c)\, \delta c_i\\
  &=\sum_i\Big[\pd{\E}{c_i}\big(\c,\x^{(n)}(\c)\big)\ + \nonumber\\
  &\hspace{5em} \sum_k \pd{\E}{x_k} \big(\c,\x^{(n)}(\c)\big)
                    \  \pd{x^{(n)}_k}{c_i}(\c) \Big]\, \delta c_i \nonumber\ .
\end{align}
Due to the stationarity condition \eqref{eq:stationary} the second part
in the square brackets vanishes so that we obtain for stationary solutions
\begin{align}\label{eq:HFgeneral}
\pp{c_i}\ E_n(\c) =
\frac{\partial \E}{\partial c_i}&\big(\c,\x^{(n)}(\c)\big)\ .
\end{align}
The derivative of the energy at the stationary solutions is just
the partial derivative of the energy functional with respect
to the external parameter calculated with the stationary state.
This generalizes the  Hellmann--Feynman theorem \eqref{eq:HF}.

Taking
\begin{equation}
\E(\c,\x)=\frac{\matrixe{\x}{H(\c)}{\x}}{\braket{\x}{\x}}\ \ \mbox{and}\ \
  \ket{\x}=\sum_k x_k \ket{k}\ ,
\end{equation}
where $\ket{k}$ denotes some fixed basis and doing a variation with
respect to the parameters $\x=\{x_1,x_2,\dots\}$ yields the
time-independent Schr\"odinger equation \eqref{eq:Schroedinger} and the original
Hellmann--Feynman theorem \eqref{eq:HF} as a special case of \eqref{eq:HFgeneral}.

It is interesting to note that the Hellmann--Feynman theorem even applies to
every solution $\ket{\Psi(t)}$ of the time-dependent Schr\"odinger equation
\begin{equation}\label{eq:TimedepSchroedinger}
i\ddt\ket{\Psi(t)}=H\big(\c(t)\big)\ \ket{\Psi(t)}\ .
\end{equation}
For the total time derivative of the mean energy, which is the expectation value
of the Hamiltonian, one obtains
\begin{align}
\frac{d E}{d t}=&\ddt\frac{\matrixe{\Psi(t)}{H\big(\c(t)\big)}{\Psi(t)}}
                       {\braket{\Psi(t)}{\Psi(t)}}\nonumber\\
      =&\sum_i\frac{\matrixe{\Psi(t)}{\pd{}{c_i}H\big(\c(t)\big)}{\Psi(t)}}
                       {\braket{\Psi(t)}{\Psi(t)}}\
                       \frac{d\,c_i}{dt}(t)\nonumber\\
  &+\left[\frac{\matrixe{\Psi(t)}{H\big(\c(t)\big)\ddt}{\Psi(t)}}
                       {\braket{\Psi(t)}{\Psi(t)}}\ +\ h.a.\right]\ .
\end{align}
Here we used the fact that a hermitian Hamiltonian does not change
the norm of the state $\ket{\Psi(t)}$.
Inserting the time-dependent Schr\"odinger equation
\eqref{eq:TimedepSchroedinger} into the expression in the
square brackets lets this term of the equation vanish so that
the time-dependent Hellmann--Feynman theorem reads:
\begin{align}\label{eq:HFtime}
\frac{d E}{d t}=&\sum_i\frac{\matrixe{\Psi(t)}{\pd{}{c_i}H\big(\c(t)\big)}{\Psi(t)}}
                       {\braket{\Psi(t)}{\Psi(t)}}\  \frac{d\,c_i}{dt}(t)\ .
\end{align}
Again the variation of the energy is given by the expectation value of the
derivatives of the Hamiltonian with respect to the external parameters.
See also Ref. \cite{DiVP00}.

The generalized theorem \eqref{eq:HFgeneral} is even valid for
classical mechanics, just take $\x$ to be the canonical variables
$p$ and $q$. The equations of motion are
\begin{align}
\frac{d\,p}{dt}=-\pd{\E}{q}(\c,p,q)\ \ \mbox{and}\ \
\frac{d\,q}{dt}=\pd{\E}{p}(\c,p,q) \ .
\end{align}
At the stationary points the partial
derivatives of the Hamilton function $\E(\c,p,q)$ vanish and one
obtains again Eq. \eqref{eq:HFgeneral}.


%
\section{Models}\label{sec:models}
In this section we discuss briefly the Hartee-Fock (HF) method when a Hamiltonian
is used, the extension to HF with density-matrix functionals and
the inclusion of pairing correlations. We show that in all cases the
generalized Hellmann--Feynman relation holds. We also explain in short the
quantities discussed in the section containing calculations for
$^{229}$Th.

\subsection{Hartree-Fock with Hamiltonian}
In the HF approximation one uses a single Slater determinant
\begin{equation}
\ket{\Psi_\HF}=a^\dagger_1\, a^\dagger_2\, \cdots a^\dagger_A\, \ket{\emptyset}
\end{equation}
as the many-body trial state. The creation operators $a^\dagger_\nu$
that create the occupied single-particle states $\ket{\phi_\nu}$,
\begin{equation}
a^\dagger_\nu\:\ket{\emptyset}=\ket{\phi_\nu}=\sum_i c^\dagger_i\:\ket{\emptyset}\:D_{i\nu}\ ,
\end{equation}
are represented in a working basis $\ket{i}=c^\dagger_i\:\ket{\emptyset}$.
Thus, the expansion coefficients $D_{i\nu}$ represent the set
$\bs{x}=\{D_{i\nu};\nu=1,\dots,A,i=1,2,\dots\}$ of variational parameters.

The general definition of the one-body density operator
\begin{equation}
\hat{\rho}=\sum_{i,k}\ket{i}\:\rho_{ik}\: \bra{k}
\end{equation}
is given in terms of the expectation values of $c^\dagger_k c_i$, where
the creation operators $c^\dagger_i$ create the single-particle basis $\ket{i}$.
In the HF case:
\begin{equation}\label{eq:densmat}
\rho_{ik}=\frac{\matrixe{\Psi_\HF}{c^\dagger_k c_i}{\Psi_\HF}}
                {\braket{\Psi_\HF}{\Psi_\HF}}
         =\frac{\sum_\nu D_{i\nu} (D_{k\nu})^*}{\braket{\Psi_\HF}{\Psi_\HF}}
               \ .
\end{equation}
%
The energy of the HF Slater determinant can be expressed
in terms of the idempotent ($\hat{\rho}^2=\hat{\rho}$) one-body density as
\begin{align}\label{eq:HFenergy}
{\cal E}_\HF[\,\c,\hat{\rho}\,]=
           \frac{\matrixe{\Psi_\HF}{H(\c)}{\Psi_\HF}}
                {\braket{\Psi_\HF}{\Psi_\HF}}&\nonumber \\
    =\sum_{ij} t_{ij}(\bs{c})\: \rho_{ji}
              + \frac{1}{4} \sum_{ijkl} v_{ik,jl}(\bs{c})&\:\rho_{ji}\rho_{lk}\nonumber \\
  + \frac{1}{36} \sum_{ijklmn} v_{ikm,jln}(\bs{c})&\:\rho_{ji}\rho_{lk}\rho_{nm}+\cdots
  \ ,
\end{align}
where $t_{ij}(\c)$ denotes the matrix elements of the kinetic energy,
$v_{ij,kl}(\c)$ the antisymmetrized matrix elements of the two-body interaction,
$v_{ikm,jln}(\c)$ of the three-body interaction, and so on. The dependence
on the parameter set $\c$ that includes the
nucleon masses, coupling strengths, interaction ranges etc. will not be
indicated again until required.
The variational parameters $\x=\{D_{i\nu}\}$ reside in the one-body density matrix $\hat{\rho}$
as given in Eq.~\eqref{eq:densmat}.
In order to work with familiar expressions in this section we will write
$\hat{\rho}$ instead of $\x$.

Variation of the energy given in Eq.~\eqref{eq:HFenergy}
with respect to $\hat{\rho}$ leads to the HF equations
\begin{equation}\label{eq:HartreeFock}
\hat{h}_\mathrm{HF}[\hat{\rho}]\ \hat{\rho} = \hat{\rho}\ \hat{h}_\mathrm{HF}[\hat{\rho}]\ .
\end{equation}
The one-body HF Hamiltonian
\begin{equation}
\hat{h}_\HF[\,\hat{\rho}\,]=\sum_{i,k}\ket{i}\:h_\HF[\,\hat{\rho}\,]_{ik}\:\bra{k}
\end{equation}
is a functional of the one-body density and its matrix elements are given by
\begin{equation}\label{eq:HFhamiltonian}
h_\HF[\,\hat{\rho}\,]_{ik}=
       \pp{\rho_{ki}}\, {\cal E}_\HF[\,\c,\hat{\rho}\,]\ ,
\end{equation}
or in short notation
\begin{equation}\label{eq:HFHamiltonian}
\hat{h}_\HF[\,\hat{\rho}\,]=
       \frac{\delta}{\delta\hat{\rho}}\, {\cal E}_\HF[\,\c,\hat{\rho}\,]\ .
\end{equation}
The eigenstates $\ket{\phi_\nu}$
of $\hat{h}_\HF[\,\hat{\rho}\,]$ represent the basis in which
both, $\hat{h}_\HF[\,\hat{\rho}\,]$ and $\hat{\rho}$ are diagonal:
\begin{align}\label{eq:H_HF}
\hat{h}_\HF[\,\hat{\rho}\,]&=\sum_{\nu} \ket{\phi_\nu}\ \epsilon_\nu \ \bra{\phi_\nu}\\
\hat{\rho}&=\sum_{\nu} \ket{\phi_\nu}\ n_\nu\ \bra{\phi_\nu} \ ,
\end{align}
where $\epsilon_\nu$ denotes the single-particle energy and
$n_\nu$ are the single-particle occupation numbers
that are zero or one in the case of a single Slater determinant.

Eq.~\eqref{eq:HartreeFock} represents the stationarity condition
\eqref{eq:stationary} and hence the HF approximation
fulfills the Hellmann--Feynman theorem.

In nuclear structure theory the microscopic nucleon-nucleon interaction
induces strong short-range correlations that cannot be represented
by a single Slater determinant. Therefore the HF method
as explained here cannot be used. For example the strong short-ranged
repulsion makes all two-body matrix elements $v_{ik,lm}$ positive and
large, so that the HF Slater determinant does not give
bound objects. The way out is to use effective interactions that
incorporate the short-range correlations explicitly,
see for example Ref.~\cite{RPPH06,RNHF04,NefF03}.
Another approach is the density-matrix functional theory which we discuss
in the following section.

\subsection{Hartree-Fock with density-matrix functionals}
It has turned out that bypassing the construction of an effective microscopic
Hamiltonian by postulating an ansatz for the energy as functional of the one-body
density-matrix $\hat{\rho}$, as originally proposed by Skyrme for the non-relativistic
and by J.~Boguta and A.R.~Bodmer \cite{BogB77} and D.~Walecka \cite{Wale74}
for relativistic nuclear physics (or by Kohn and Sham \cite{KohS65} for
the atomic case), is very successful in describing ground state properties.
The energy functional ${\cal E}_\DF[\,\c,\hat{\rho}\,]$ contains
a finite number of parameters, $\c$, that are adjusted by fitting observables
to nuclear data.
The shape of the functional is subject of past and present research
\cite{EBGK75,Ring96,BeHR03}
and is being improved to also apply for nuclei far off stability.
Different from the HF case with Hamiltonian, Eq.~\eqref{eq:HFenergy}, the
densities may appear also with non-integer powers and the exchange
terms are not calculated explicitly but absorbed in the form of the energy functional.

Not all of the information residing in the one-body density-matrix $\hat{\rho}$
is used. Usually one uses the proton and neutron density
$\rho_p(\vek{r}),\rho_n(\vek{r})$,
kinetic energy densities $\tau_p(\vek{r}),\tau_n(\vek{r})$,
current densities $\vek{j}(\vek{r})$, etc.
\begin{align}
\E_\DF[\c,\hat{\rho}]=\E_\DF(\c,\rho_p(\vek{r}),\rho_n(\vek{r}),\tau_p(\vek{r}),\tau_n(\vek{r}),
               \vek{j}(\vek{r}),\dots)
\end{align}
In order to keep densities and currents consistent and corresponding to fermions they are
expressed in terms of the single-particle states $\ket{\phi_\nu}$
\begin{align}\label{eq:sp-state}
\ket{\phi_\nu}=\sum_i \  \ket{i}\: D_{i\nu}
\end{align}
that represent the occupied states of a single Slater determinant.
As they are expanded in terms of a working basis $\ket{i}$
the energy $\E_\DF[\,\c,\hat{\rho}\,]$ is, like in the HF case, a function of the
variational parameters $\x=\{D_{i\nu};\ i,\nu=1,2,\dots\}$ or $\x=\hat{\rho}$.

The difference to HF with a Hamiltonian is that the mean-field
Hamiltonian $\hat{h}_\MF[\,\hat{\rho}\,]$ obtained by the functional derivative
of the density-matrix functional $\E_\DF$
\begin{equation}\label{eq:MFHamiltonian}
\hat{h}_\MF[\,\hat{\rho}\,]=
       \frac{\delta}{\delta {\hat{\rho}}} {\cal E}_\DF[\,\c,\hat{\rho}\,]
\end{equation}
is not given by a microscopic Hamiltonian anymore, but by
the functional form of $\E_\DF$ and the fitted parameters in the set $\c$.

The stationarity conditions \eqref{eq:stationary} lead to
the self-consistent mean-field equations
\begin{equation}\label{eq:meanfield}
\hat{h}_\mathrm{MF}[\hat{\rho}]\ \hat{\rho} = \hat{\rho}\ \hat{h}_\mathrm{MF}[\hat{\rho}]
\end{equation}
that have the same structure as the HF equations.

Because the self-consistent solution is obtained by searching for
solutions of the stationarity conditions \eqref{eq:stationary}.
the Hellmann--Feynman theorem
\eqref{eq:HFgeneral} is fulfilled, even if one cannot refer to a
microscopic Hamiltonian and a many-body state anymore.

One should note that it is not mandatory that the single-particle
states $\ket{\phi_\nu}$ with lowest single-particle energies are
occupied. Any combination of occupied states leads to a stationary
solution fulfilling Eq.~\eqref{eq:meanfield}.

Another interesting case is a one-body density with fractional occupation
numbers $0 \le n_\nu \le 1$ that may also commute with the mean-field Hamiltonian
and hence fulfills the stationarity condition \eqref{eq:meanfield}
so that the Hellmann--Feynman theorem is applicable.
In such a situation one cannot attribute a single Slater determinant to
the one-body density anymore, because $\hat{\rho}\neq \hat{\rho}^2$ is not
idempotent.

In any case the occupation numbers play the role of external parameters
and should be regarded as members of the set $\c$ and not as variational
parameters.

\subsection{Hartree-Fock-Bogoliubov}
The solution of the eigenvalue problem of $\hat{h}_\MF[\,\hat{\rho}\,]$
provides not only the coefficients $D_{i\nu}$ of  the occupied single-particle states
but also a representation for empty states so that one has a complete
representation of the one-body Hilbert space. With that one can
define creation operators for fermions for occupied and empty states
\begin{equation}
a^\dagger_\nu  =  \sum_i\: c^\dagger_i \: D_{i\nu} \ ,
\end{equation}
with
\begin{equation}
\ket{\phi_\nu} =a^\dagger_\nu\: \ket{\emptyset}   \ \ \mathrm{and}\ \
\ket{i}    = c ^\dagger_i\: \ket{\emptyset} \ .
\end{equation}
and their corresponding annihilation operators $a_\nu$ and $c_i$.

Pairing correlations in the many-body state can be incorporated by
Bogoliubov quasi-particles that are created by
\begin{equation}\begin{array}{rl}
\alpha^\dagger_\nu       &= u_\nu a^\dagger_\nu - v_\nu a_{\bar\nu}\\
\alpha^\dagger_{\bar\nu} &= u_\nu a^\dagger_{\bar\nu} + v_\nu a_\nu
\end{array}\end{equation}
as linear combinations of the creation and annihilation operators,
$a^\dagger_\nu, a_{\bar{\nu}}$,
of the eigenstates of the
one-body density matrix (the so-called canonical states).
The parameters $u_\nu$ and $v_\nu$ can be chosen real and the requirement
that $\alpha^\dagger_\nu$ and $\alpha^\dagger_{\bar\nu}$ are fermionic
quasi-particle operators implies $u_\nu^2+v_\nu^2=1$.
The pairing partner states $\nu$ and $\bar{\nu}$ are usually
mutually time-reversed states.

The many-body trial state is expressed as
\begin{equation}\label{eq:HFBstate}
\ket{\Psi_{\rm{HFB}}}=\prod_\mu \: a^\dagger_\mu \: \prod_\nu \left(\sqrt{1-v_\nu^2}
    + v_\nu\; a^\dagger_\nu a^\dagger_{\bar\nu}\right)\: \ket{\emptyset} \ ,
\end{equation}
where the product over $\mu$ runs over the so called blocked states
or unpaired states and $\nu$ runs over all other paired states.
Besides the variational parameters residing in the operators
$a^\dagger_\nu$ that create eigenstates of the mean-field Hamiltonian
the energy depends now also on the variational parameters  $v_\nu$,
hence $\x=\{D_{i\nu},v_\nu;\ i,\nu=1,2,\dots\}$.

As the trial state \eqref{eq:HFBstate} has no sharp particle number
the stationarity conditions Eq.~\eqref{eq:stationary} have to be augmented
by a constraint on mean proton number ${\cal Z}$ and mean neutron number
${\cal N}$ to obtain the selfconsistent
Hartree-Fock-Bogoliubov (HFB) equations:
\begin{equation}\label{eq:HFBfunctional}
\E_\rm{HFB}=\E-\lambda_p {\cal Z} -\lambda_n {\cal N} \ .
\end{equation}
The proton and neutron chemical potentials, $\lambda_p$ and $\lambda_n$,
have to be regarded as members of the set $\bs{c}$ of external parameters.
The additional constraints do not alter the arguments leading the
Hellmann--Feynman theorem, thus it is also valid in the HFB case.

In HFB it is convenient to introduce a generalized density matrix
\begin{equation}
\R=\left( \begin{array}{cc}
                \hat{\rho}   & \hat{\kappa}\\
              -\hat{\kappa}^*& 1-\hat{\rho}^*
                                \end{array} \right)
\end{equation}
where $\hat{\rho}$ is the normal one-body density
\begin{equation}
\rho_{ik}=\frac{\matrixe{\Psi_\rm{HFB}}{c^\dagger_{k} c_i}{\Psi_\rm{HFB}}}
                    {\braket{\Psi_\rm{HFB}}{\Psi_\rm{HFB}}}
\end{equation}
and $\kappa$ the so called abnormal density
\begin{equation}
\kappa_{ik}=\frac{\matrixe{\Psi_\rm{HFB}}{c_k c_i}{\Psi_\rm{HFB}}}
                      {\braket{\Psi_\rm{HFB}}{\Psi_\rm{HFB}}} \ .
\end{equation}
Both can be expressed in terms of $D_{i\nu}$ and $v_\nu$.
The generalized density matrix is idempotent and hermitian
\begin{equation}
\R^2=\R\ \ \rm{and} \ \ \R^\dagger=\R
\end{equation}
and the stationarity condition leads to
\begin{equation}
\HM\!\big[\R\big]\: \R=
                      \R\:\HM\!\big[\R\big]
\end{equation}
quite in analogy to the mean-field equations \eqref{eq:meanfield} without
pairing correlations. The pseudo-Hamiltonian
\begin{align}
\HM\!\big[\R\big]&=
       \frac{\delta}{\delta \R}\, \E_\rm{HFB}\big[\bs{c},\R\big]\nonumber\\
       &= \left( \begin{array}{cc}
                \hat{h}_\rm{MF}-\lambda   & \hat{\Delta}\\
              -\hat{\Delta}^*& \ \ \lambda-\hat{h}_\rm{MF}^*
                                \end{array} \right)
\end{align}
contains the mean-field Hamiltonian and the pairing part $\hat{\Delta}$.
The chemical potentials $\lambda=(\lambda_p,\lambda_n)$ determine
the mean proton and neutron number.
For details and further reading see Refs.~\cite{RinS80, DNWB96, KSTV06}.

Let us write down only the one-body density  because it is referred to in
the results for $^{229}$Th.
The one-body density matrix is given by
\begin{equation}
\hat{\rho}=\sum_{\nu\ \rm{paired}}\ket{\phi_\nu}\, v_\nu^2\, \bra{\phi_\nu}
     +\sum_{\mu\ \rm{blocked}} \ket{\phi_\mu}\, n_\mu \, \bra{\phi_\mu} \ ,
     \end{equation}
where the $\ket{\phi_\nu}$ denote canonical basis states.
The occupation numbers are given by $n_\nu=n_{\bar\nu}=v_\nu^2$ for the paired states
and are the same for $\nu$ and $\bar\nu$. For the unpaired or blocked
states $n_\mu=0,1$. In the application we will consider only one blocked
neutron state for $^{229}$Th.

%
\section{Amplification}\label{sec:amplification}

To be more specific let us consider as external parameters the
fine structure constant $\alpha$ and the proton and neutron mass $m_p, m_n$,
thus $\c=\{\alpha,m_p,m_n\}$.
In this Section, we write down expressions pertaining to a Hamiltonian;
those characteristic to a density functional are entirely analogous,
cf.~Sec.~\ref{sec:HellmannFeynman}.
The partial derivatives of a non-relativistic Hamiltonian
are
\begin{align}
\pd{}{\alpha}H(\alpha,m_p,m_n)&=\frac{1}{\alpha}\ V_C\\
\pd{}{m_p} H(\alpha,m_p,m_n)&=Z - \frac{1}{m_p}\ T_p\\
\pd{}{m_n} H(\alpha,m_p,m_n)&=N - \frac{1}{m_n}\ T_n \ ,
\end{align}
where $V_C$ denotes the operator for the Coulomb energy,
$Z,N$ the proton and neutron number, respectively,
and $T_p,T_n$ stand for the proton and neutron kinetic energy operator,
respectively.

According to the Hellmann--Feynman theorem a small variation
$\delta \alpha$ of the fine structure constant
results in a variation of an energy eigenvalue given by
\begin{align}\label{eq:alphavariation}
\delta &E_n = \bigg(\matrixe{\Psi_n}{V_C}{\Psi_n}
                     +Z\,\alpha\dd{m_p}{\alpha} + N\,\alpha\dd{m_n}{\alpha} \\
                 &-\matrixe{\Psi_n}{T_p}{\Psi_n}\frac{\alpha\dd{m_p}{\alpha}}{m_p}
                  -\matrixe{\Psi_n}{T_n}{\Psi_n}\frac{\alpha\dd{m_n}{\alpha}}{m_n}
              \bigg)\frac{\delta \alpha}{\alpha}\ .\nonumber
\end{align}
In principle there could also be a dependence of the nuclear interaction
$V_N$ on $\alpha$, e.q., through meson masses, which we neglect here.
The dependence of the nucleon masses on $\alpha$ can be estimated from
the paper by Mei{\ss}ner et al. \cite{MRWY07} (Eq. (36)). Their estimate of the
neutron-proton mass difference due to the electromagnetic interaction is
$\Delta m^{\rm{(EM)}}_{np}=-0.68$~MeV which yields
\begin{align}
\frac{\alpha\dd{m_p}{\alpha}}{m_p}=- \frac{\Delta m^{\rm{(EM)}}_{np}}{2m_p}
                           &\approx +0.36\cdot 10^{-3}\\
\frac{\alpha\dd{m_n}{\alpha}}{m_n}=\ \frac{\Delta m^{\rm{(EM)}}_{np}}{2m_n}
                           &\approx -0.36\cdot 10^{-3}\ .
\end{align}

A possible variation of some QCD constant $c$ would lead to
\begin{align}\label{eq:varQCD}
\delta E_n = &\bigg(\matrixe{\Psi_n}{\dd{}{c} V_N(c)}{\Psi_n}
                                +Z\,\dd{m_p}{c} + N\,\dd{m_n}{c}\\
                - &\matrixe{\Psi_n}{T_p}{\Psi_n}\frac{\dd{m_p}{c}}{m_p}
                -  \matrixe{\Psi_n}{T_n}{\Psi_n}\frac{\dd{m_n}{c}}{m_n}
              \bigg)\delta c\ ,\nonumber
\end{align}
where $V_N$ is the nuclear part of the interaction. This variation is
actually linked to the dimensionless ratio $c=m_q/\Lambda_{QCD}$, where
$m_q$ denotes the current quark mass and $\Lambda_{QCD}$ the strong interaction scale
\cite{FlaS02,FlaS03,Flam06}.

While the dependence of the nucleon mass on the current quark mass
has been calculated \cite{FHJR06,FLTY04}, the QCD constants
enter the effective interaction $V_N(c)$ in a very complicated and yet unknown way.
In this paper we do not consider such variations of QCD constants explicitly
but calculate besides the total energies the kinetic energies
and the Coulomb energies which then can be combined according to
Eqs. \eqref{eq:alphavariation} or \eqref{eq:varQCD} to obtain the variations
with respect to variations of $\alpha$ or QCD parameters.

As the temporal variation of the fundamental constants $\alpha, m_p$ and $m_n$ are tiny,
if they exist at all, it has been proposed to consider
transition frequencies
\begin{align}\omega&=E_1(\alpha)-E_0(\alpha)\end{align}
that can be measured with high precision.

If the two energy levels belong to the same nucleus the terms with
the proton and neutron number drop out and we obtain for the
relative variation $\delta\omega/\omega$ of the transition frequency
\begin{align}\label{eq:amplify}
\frac{\delta \omega}{\omega}& = \frac{1}{\omega}
                      \bigg(\pd{E_1}{\alpha}-\pd{E_0}{\alpha}\bigg)\delta \alpha\\
&=\frac{1}{\omega}
  \bigg(\Delta V_C\!-\!\Delta T_p \frac{\alpha\dd{m_p}{\alpha}}{m_p}
 \!-\!\Delta T_n \frac{\alpha\dd{m_n}{\alpha}}{m_n} \bigg)\nonumber
 \frac{\delta \alpha}{\alpha}\\
 &=A\;\frac{\delta \alpha}{\alpha}\nonumber
\end{align}
with the abbreviation
\begin{align}
\Delta X = \frac{\matrixe{\Psi_1;\alpha}{X}{\Psi_1;\alpha}}
                           {\braket{\Psi_1;\alpha}{\Psi_1;\alpha}}
    -               \frac{\matrixe{\Psi_0;\alpha}{X}{\Psi_0;\alpha}}
                           {\braket{\Psi_0;\alpha}{\Psi_0;\alpha}}
\end{align}
for the difference of the expectation values of the operators
$X=\{V_C,T_p,T_n\}$ calculated with the stationary states.

The results discussed in Sec.~\ref{sec:results} show that
$\Delta T_n$ and $\Delta T_p$ are of the same order as $\Delta V_C$
so that the terms with the proton and neutron mass variations can be neglected.

If there is a system, where the energy difference $\omega$ is much smaller
than the difference between the Coulomb energies of the two states
one gets an amplification factor in the measurement of $\delta \alpha/\alpha$.
Flambaum \cite{Flam06} has proposed the nucleus $^{229}$Th as a good
candidate \cite{PeiT03} because it possesses two almost degenerate eigenstates with
$\omega\approx$~8~eV   according to a recent measurement \cite{BBBB07}.

The theoretical task is to investigate these states carefully in order to get
a reliable estimate for their Coulomb and kinetic energies.
For the calculation of  these quantities we employ in the following section
state-of-the-art mean-field models
and also include the effects of pairing correlations.

\section{Results for $\mathbf{^{229}}\rm{\mathbf{Th}}$} \label{sec:results}

\begin{figure}[t]
\begin{center}
\includegraphics*[scale=.8]{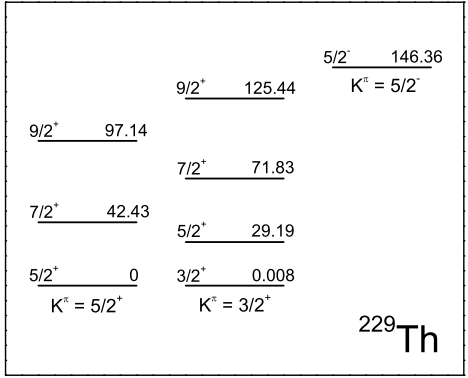}
\end{center}
\caption{\label{fig:0}Measured low lying states of $^{229}$Th with spin and
parity assignments \cite{AuWT03}.
Energies are in keV.}
\end{figure}

The nucleus $^{229}$Th with 90 protons and 139 neutrons occurs in nature as
the daughter of the $\alpha$-decaying $^{233}$U and decays itself with a half life of
7880 years, again by $\alpha$ emission.
This nucleus has attracted lot of interest as it has the lowest lying excited state known.
In Fig.~\ref{fig:0} the spectrum is arranged in terms of rotational bands
\cite{AuWT03}.
Two low lying rotational bands with $K^\pi=5/2^+$ and $3/2^+$
can be identified, with band heads that according to recent measurements
differ in energy by only about 8~eV. The first negative parity band
with $K^\pi=5/2^-$ occurs at 146.36~keV.

Because of the amplification effect discussed in Sec.~\ref{sec:amplification}
the transition $3/2^+ \rightarrow 5/2^+$ is regarded
as a possible candidate to measure the time variation of the fine structure constant.
A simple estimate of the moment of inertia for the two bands
assuming a $J(J+1)$ energy dependence shows that the two rotational bands have
very similar intrinsic deformation. Therefore, one does not expect a large
difference between the Coulomb energies of the $5/2^+$ and $3/2^+$ band heads that
would be due to differences in shapes. Instead, one has to consider and
work out detailed effects due to different configurations of these states.
%

\begin{figure*}[t]
\begin{center}
\includegraphics*[scale=.9]{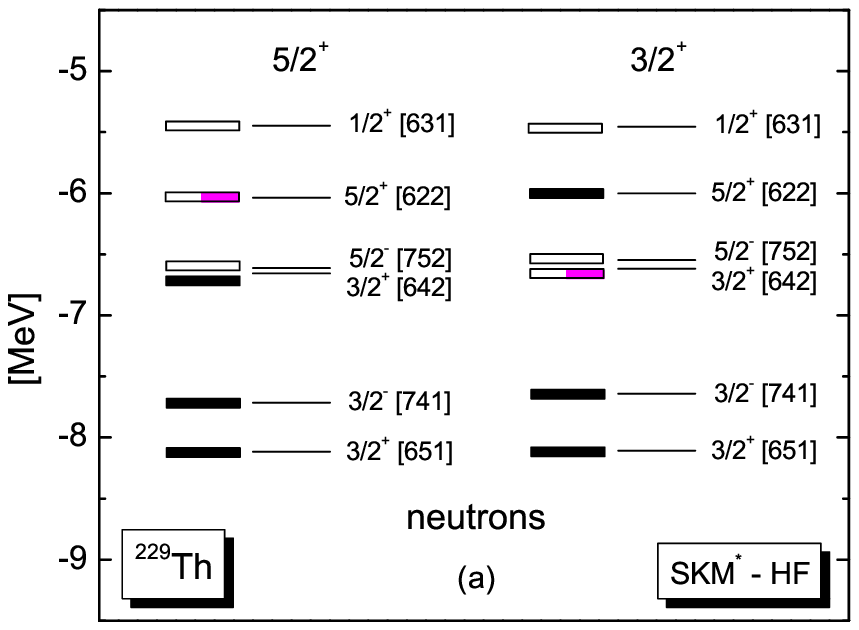}
\includegraphics*[scale=.9]{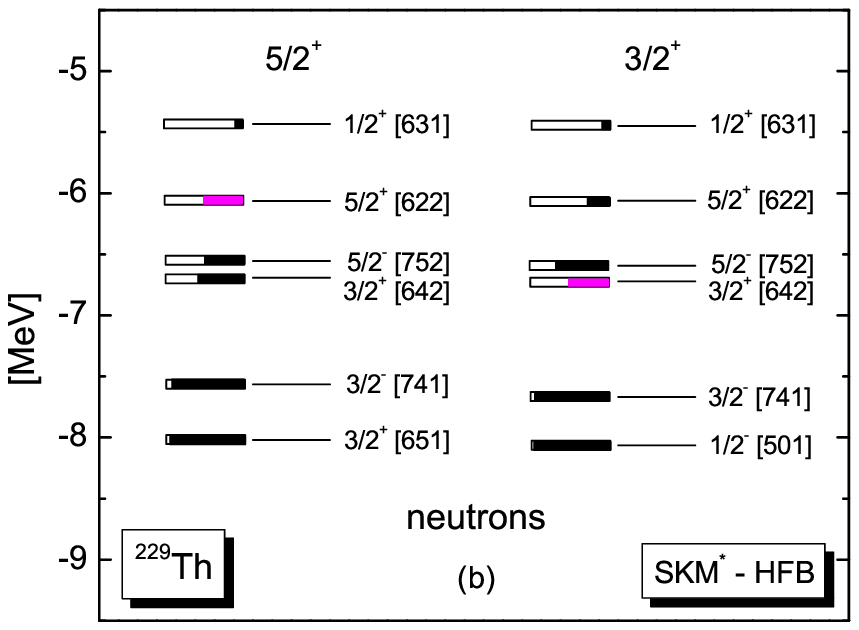}
\end{center}
\caption{\label{fig:1}(Color online) Neutron single-particle energies and occupation numbers
labeled by the asymptotic Nilsson quantum numbers
$\nu=\Omega^\pi[N,N_z,\Lambda]$ for the SkM$^*$ energy functional.
(a) HF mean-field energies $\epsilon_\nu$
and occupation numbers $n_\nu$.
Full bars denote $n_\nu+n_{\bar\nu}=2$, i.e. two particles in degenerate pair
of states with $m_j=\pm\Omega$.
Half full gray (pink) bars denote one particle, $n_\mu=1,\:n_{\bar\mu}=0$.
(b) HFB 
{mean-field} 
energies $\epsilon_\nu$ 
{(eigenvalues of $\hat{h}_\MF$)}
and occupation numbers
{$v^2_\nu$}.
Length of bars indicates $n_\nu+n_{\bar\nu}$. Gray (pink) bars stand for blocked states
with $n_\mu=1,\:n_{\bar\mu}=0$. 
}
\begin{center}
\includegraphics*[scale=.9]{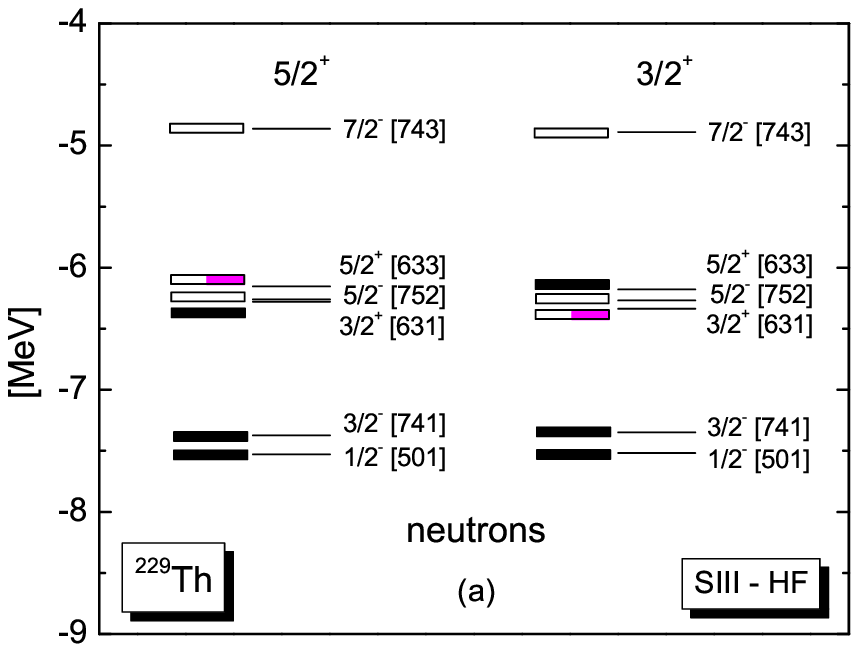}
\includegraphics*[scale=.9]{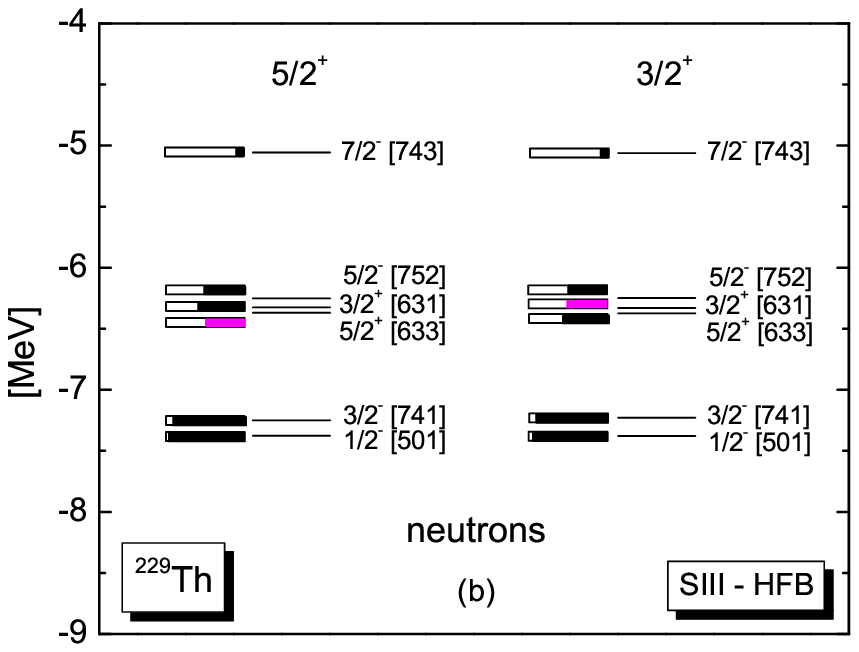}
\end{center}
\caption{\label{fig:2}(Color online) Same as Fig.~\ref{fig:1}, but for the SIII energy functional.}
\end{figure*}
%
\begin{table*}[tb]
\caption{Total, Coulomb, neutron and proton kinetic energies of the
$^{229}$Th $5/2^+$ ground state calculated with different energy functionals.
Differences of these energies between $3/2^+$ first excited state  and $5/2^+$ ground state.
\label{tab:energies}}
\begin{center}
\vspace{6mm}
\tabcolsep=1.em \renewcommand{\arraystretch}{1.3}%
\begin{tabular}
[c]{r|c|c|c|c|c|c} \hline\hline
                & Exp.   &\multicolumn{2}{|c}{SkM$^*$} & \multicolumn{2}{|c|}{SIII} & NL3 \\
\hline
  $5/2^+$&Ref.~\cite{AuWT03}&    HF    &     HFB   &     HF    & HFB       &    RH     \\
\hline
$E^{tot}$ [MeV]&-1748.334&-1739.454 & -1747.546 & -1741.885 & -1748.016 & -1745.775 \\
$V_C$ [MeV]     &        &  923.927 &   924.854 &   912.204 &   912.216 &   948.203 \\
$T_n$ [MeV]     &        & 2785.404 &  2800.225 &  2783.593 &  2794.909 &  2059.640 \\
$T_p$ [MeV]     &        & 1458.103 &  1512.705 &  1442.018 &  1477.485 &  1106.697 \\
\hline
$3/2^+ - 5/2^+$ &Ref.~\cite{BBBB07} &           &           &           &           &     \\
\hline
$\Delta E^{tot}$[MeV]&0.000 008 & 0.619  & -0.046 &  0.141 & -0.074 &  2.407 \\
$\Delta V_C$ [MeV]     &        & 0.451  & -0.307 & -0.098 &  0.001 &  1.011 \\
$\Delta T_n$ [MeV]     &        & 2.570  &  0.954 & -0.728 &  0.087 & -2.181 \\
$\Delta T_p$ [MeV]     &        & 0.688  &  0.233 & -0.163 & -0.022 & -1.996 \\
\hline
\hline
\end{tabular}
\end{center}
\end{table*}

\subsection{Hartree-Fock with density-matrix functional theory}

To come to more quantitative statements we calculate the energies
of these two states with the best methods available:
density-matrix functional theory without and with pairing, and a
spherical relativistic mean-field theory \cite{GaRT90}. In the
non-relativistic case we use the computer program HFODD (v2.33j)
\cite{[Dob04fw],[Dob07w]} and employ two successful energy
functionals, SIII \cite{SIII75} and SkM$^*$ \cite{SKM*82}. Because
of large uncertainties related to polarization effects due
time-odd mean fields \cite{[Zal08]}, in the present study these
terms in the energy functionals are neglected. The standard Slater
approximation is used to calculate the Coulomb exchange energies.
The code works with a Cartesian harmonic oscillator (HO)
eigenstates as working basis $\ket{i}$ and allows for triaxial and
parity-breaking deformations, but the states determined in
$^{229}$Th turn out to have axial shapes with conserved parity. In
our calculations we have used the basis of HO states up to the
principal quantum number of $N_0=18$ and the same HO frequency of
$\hbar\omega=8.05$\,MeV in all three Cartesian directions.

In the discussion, the resulting basis $\ket{\phi_\nu}$
will be assigned Nilsson quantum numbers $\Omega^\pi[N,N_z,\Lambda]$
(for details see Ref.~\cite{RinS80,BohrMottelson}) by looking for the largest overlap
$\left|\braket{\Omega^\pi[N,N_z,\Lambda]}{\phi_\nu}\right|$
with a Nilsson state.
$N=N_x+N_y+N_z=N_\rho+N_z+\Lambda$ denotes the total number of
oscillator quanta, $N_z$ the number of quanta in the direction of the symmetry axis.
$\Lambda=N_\rho, N_\rho -2, \dots 0\; \mathrm{or}\; 1$ and $\Omega=\Lambda\pm\half$
are the absolute values the projection
of orbital angular momentum and total spin on the symmetry axis, respectively.
$\pi=(-1)^N$ is the parity.

It turns out that for both functionals the lowest HF state has $\Omega^\pi=5/2^-$
and thus corresponds to the intrinsic state of the $K^\pi=5/2^-$ band
that is experimentally located at 146.36~keV.
The total binding energy amounts to $-1739.454$~MeV for SkM$^*$ and $-1741.885$~MeV for SIII
which should be compared to the experimental energy of $-1748.334$~MeV \cite{AuWT03}.
On this absolute scale the calculated HF energies are already
amazingly good.

In order to get the experimentally observed parity
we perform the variation procedure in the subspace with positive parity.
For both energy functionals the Slater determinant with the lowest energy
has $K^\pi=5/2^+$.
All energies are summarized in Table~\ref{tab:energies}.
The last neutron occupies the level labeled with 5/2$^+$[622] in the SkM$^*$
and with 5/2$^+$[633] in the SIII case.
One should keep in mind that the single-particle states are superpositions of
Nilsson states and the labeling refers only to the largest component.
The states are detailed in Eqs.~\eqref{eq:Nilsson-SKM-HF}-\eqref{eq:Nilsson-SIII-HFB} below.

The neutron single-particle energies $\epsilon_\nu$ and the occupation numbers
are displayed in Fig.~\ref{fig:1}(a) for SkM$^*$ and in Fig.~\ref{fig:2}(a) for SIII.
The 5/2$^-$[752] state which would be occupied in the negative parity case
is very close to the 5/2$^+$[633] state -- for SIII they are almost degenerate.

As can be seen from Table~\ref{tab:energies}, the total HF binding energy agrees
with the measured one up to about 9~MeV for the SkM$^*$ and up to about 6~MeV for
the SIII energy functional. Keeping in mind that no parameters have been
adjusted to the specific nucleus considered here it is surprising that these
mean-field models can predict the energy with an uncertainty of only about 0.5~\%.

Putting two neutrons in the 5/2$^+$[622] (for SkM$^*$) or 5/2$^+$[633] (for SIII) level
and one in the 3/2$^+$[642] (for SkM$^*$) or 3/2$^+$[631] (for SIII) level
and minimizing the total energy yields an excited HF state that
is to be regarded as the intrinsic state of the experimentally observed $K^\pi=3/2^+$ band.
As can be seen in Table~\ref{tab:energies}, the excited states occur
at 0.619~MeV for the SkM$^*$ and at 0.141~MeV for the SIII density functional.
The difference in Coulomb energies $\Delta V_C$ amounts to 0.451~MeV for
SkM$^*$ and to $-0.098$~MeV for SIII.
The kinetic energy differences $\Delta T_p$ and $\Delta T_n$ for
protons and neutrons, respectively, dissent even more.
These deviations between the two energy
functionals reflect the differences in the structure of the intrinsic states
as also seen from the difference in the single-particle states discussed above.

\subsection{Relativistic mean-field}

We also perform a spherical relativistic mean-field calculation
with the NL3 parameter set \cite{NL3-97} and find
that in the ground state the last neutron occupies a $2g_{9/2}$ orbit.
For vanishing deformation the Nilsson state 5/2$^+$[633] belongs to the
subspace spanned by the spherical $2g_{9/2}$ orbits so that this
result is not unreasonable when comparing with the deformed SIII calculation.
The leading component for the 3/2$^+$ state is for SIII
the 3/2$^+$[631] orbit (cf. Fig.~\ref{fig:2})
which for deformation zero belongs to the $1i_{11/2}$ subshell.
Therefore we create the excited state with $\Omega^\pi=3/2^+$ by a particle-hole excitation
from the $1i_{11/2}$ to the $2g_{9/2}$ shell so that there are 11 neutrons
in the $1i_{11/2}$ and 2 neutrons in the $2g_{9/2}$ shell.

The resulting energies are listed in the last column of Table~\ref{tab:energies}.
The total binding energy is similar to the non-relativistic one, but the
particle-hole excited state is 2.41~MeV higher. Different from the deformed
mean-field the single-particle states of a spherical potential have good total
spin $j$ and are $(2j+1)$-fold degenerated with large gaps between them.
The single-particle energy difference $\epsilon_{9/2}-\epsilon_{11/2}=2.74$~MeV
explains the large excitation energy of the particle-hole pair of 2.41~MeV,
which includes the rearrangement energy.

We conclude that a spherical calculation is not appropriate for this particular
question. In a recent publication \cite{HeRe08} a similar spherical relativistic
mean-field calculations comes to results comparable to ours with a Coulomb energy
difference of 0.7~MeV. Because of the unphysical properties of a spherical
$^{229}$Th we will not commit ourselves to the spherical case any longer
but proceed to consider the effects of pairing.

\begin{table*}[t]
\caption{Rms-radius and intrinsic quadrupole moments of neutron and proton densities of the
$^{229}$Th $5/2^+$ ground state calculated with different energy functionals.
Differences of these moments between $3/2^+$ first excited state  and $5/2^+$
ground state.
\label{tab:moments}}

\begin{center}
\vspace{6mm}
\tabcolsep=1.em \renewcommand{\arraystretch}{1.3}%
\begin{tabular}
[c]{r|c|c|c|c} \hline\hline
                  &\multicolumn{2}{c}{SkM$^*$} & \multicolumn{2}{|c}{SIII}    \\
\hline
$5/2^+$&   HF    &     HFB   &     HF    & HFB   \\
\hline
$R_{rms}$(neutron)[fm]    & 5.8789 & 5.8716 & 5.8971 & 5.8923 \\
$R_{rms}$(proton) [fm]    & 5.7180 & 5.7078 & 5.7817 & 5.7769 \\
$Q_{20}$ (neutron)[fm$^2$]  & 9.4407 & 9.2608 & 9.1990 & 9.0711 \\
$Q_{20}$ (proton) [fm$^2$]  & 9.5461 & 9.3717 & 9.3542 & 9.1643 \\
\hline
$3/2^+ - 5/2^+$  &   &   &   &  \\
\hline
$\Delta R_{rms}$(neutron)[fm]    & -0.0040 & 0.0036 & -0.0008 & -0.0005 \\
$\Delta R_{rms}$(proton) [fm]    & -0.0038 & 0.0039 &  0.0000 & -0.0005 \\
$\Delta Q_{20}$ (neutron)[fm$^2$]  & -0.2427 & 0.2647 & -0.0767 & -0.0516 \\
$\Delta Q_{20}$ (proton) [fm$^2$]  & -0.1824 & 0.2756 & -0.0339 & -0.0495 \\
\hline
\hline
\end{tabular}
\end{center}
\end{table*}

\subsection{Hartree-Fock-Bogoliubov}
\label{Results-pairing}

We include the pairing correlations with the Bogoliubov ansatz
\eqref{eq:HFBstate} and perform a self-consistent HFB calculation
based on the SkM$^*$ and SIII density-matrix functional. For the
SIII case proton and neutron pairing strengths of $V_0=-260$ and
$-180$\,MeV\,fm$^3$ (for a volume-type contact force) are adjusted
to reproduce the total binding energies and the odd-even
staggering with the neighboring nuclei. Proton and neutron density
matrices and pairing tensors are calculated by including
contributions from quasiparticle states up to the cutoff energy of
60\,MeV. Calculations are performed by self-consistently blocking
the  5/2$^+$ and the 3/2$^+$ quasiparticle states. In the 5/2$^+$
and 3/2$^+$ configurations, this yields the average HFB proton and
neutron pairing gaps of $\Delta_p=2.4$\,MeV and $\Delta_n=0.65$\,MeV,
and $\Delta_p=2.4$\,MeV and $\Delta_n=0.68$\,MeV, respectively, for the
SIII case and slightly larger values for the SkM$^*$ case.

The results for the energies are summarized in Table~\ref{tab:energies} and for
the mean-field single-particle energies and the occupation probabilities
in Fig.~\ref{fig:1}(b) and Fig.~\ref{fig:2}(b).
The first to note is that the excitation energy is improving. Its value decreases from
619~keV down to $-46$~keV for SkM$^*$ and from $141$~keV down to $-74$~keV for SIII.
On the accuracy level one can expect from this model the 5/2$^+$ and the 3/2$^+$
states are degenerate, like in experiment.

By looking at the occupation numbers displayed in
Figs.~\ref{fig:1}(b) and \ref{fig:2}(b) one sees that about 5 single
particle levels near the Fermi edge assume fractional occupation
numbers significantly different than 0 or 2. In the SIII case
(Fig.~\ref{fig:2}(b)) they are almost identical for the 5/2$^+$ and
3/2$^+$ states except for the 5/2$^+$[633] and 3/2$^+$[631] levels
that switch their role. For the SkM$^*$ case (Fig.~\ref{fig:1}(b))
the blocked states are energetically further away from each other
which causes more deviations in the occupation numbers.

This characteristic pattern of occupation numbers renders the HFB
results qualitatively different than the HF ones. Indeed, in the HF
case  either the 5/2$^+$ or the 3/2$^+$ orbital has the
occupation number equal to 1. Therefore, polarization effects exerted
by these two orbitals are able to render different values of
observables calculated for the 5/2$^+$ and 3/2$^+$ states. In the HFB
case, occupation numbers of the 5/2$^+$ and 3/2$^+$ orbitals are
close to 1 for both 5/2$^+$ and 3/2$^+$ configurations. Differences in
observables may here only occur due to the fact that the occupation
number of the blocked state is exactly equal to 1, while that
of the other state is approximately equal to 1, depending on
its closeness to the Fermi level.

In the paired case (HFB), we are faced with the situation, where the
zero-order approximation renders observables calculated in the
5/2$^+$ and 3/2$^+$ exactly equal. Indeed, such equality would
be the case for the 5/2$^+$ and 3/2$^+$ orbitals located exactly at
the Fermi surface and having exactly the same occupation numbers.
Note that values of all observables calculated with the HFB approach
depend only on the canonical states and occupation numbers,
irrespective of which state has been blocked in obtaining them. Of
course, one can obtain different occupation numbers for both orbitals
in question when they are energetically split. However, this may contradict the
experimental energetic degeneracy of the corresponding
configurations. All in all, within a zero-order paired approach,
polarization effects of the 5/2$^+$ and 3/2$^+$ orbitals become
exactly averaged out and the anticipated differences in observables
can occur only due to first-order corrections.

This fact is perfectly well visible in our results for the Coulomb
energy differences shown in Table~\ref{tab:energies}. For the SIII
energy functional, only 1~keV remains for $\Delta V_C$. This reduces
the amplification factor of Eq.~\eqref{eq:amplify} to about 100. For
SkM$^*$ a larger value of $\Delta V_C$ of about 300~keV is obtained
due to a larger splitting of the corresponding single-particle
orbitals. From this one must conclude that pairing correlations
result in two states with even more similar charge distributions than
in the HF calculation.

\subsection{Radii and quadrupole moments}

In Table~\ref{tab:moments} the rms-radii and quadrupole moments
{(normalized with proton and neutron number, respectively)} 
of the neutron and proton {point-}densities are
given for the ground state and as differences for the excited state.
The quadrupole moments of the protons are somewhat larger than for the neutrons.
When reoccupying the last neutron the neutron quadrupole moment decreases
substantially in the SkM$^*$-HF calculation and drags along via the nuclear interaction
the quadrupole moment of the protons. At the same time both rms-radii are decreased.
A smaller charge radius and smaller charge quadrupole moment are consistent with the increase
of the Coulomb energy which explains the large positive $\Delta V_C$ in the SkM$^*$-HF case.
When including pairing the effect goes in the opposite direction for SkM$^*$-HFB.

For the SIII functional HF and HFB calculations do not lead to noteworthy changes in
the moments when reoccupying the last neutron  and thus the Coulomb differences
remain also small. This can be anticipated when
looking at the single-particle energies of the involved neutron states
and the occupation numbers (Fig.~\ref{fig:2}).

The mean-field single-particle state that corresponds to the blocked HFB state
occupied by the unpaired neutron is
represented in Nilsson orbits for SkM$^*$-HF as
\begin{equation}\label{eq:Nilsson-SKM-HF}
\begin{array}{llll}
\ket{\frac{5}{2}^+}=&+.509\ket{622}& +.467\ket{642}& +.266\ket{862}\\
                    &+.402\ket{633}& -.397\ket{613}&+\cdots\\[1ex]
\ket{\frac{3}{2}^+}=&-.010\ket{622}& +.662\ket{642}& +.249\ket{862}\\
                    &+.305\ket{611}& -.562\ket{631}&+\cdots
\end{array}
\end{equation}
and for SkM$^*$-HFB including pairing as
\begin{equation}\label{eq:Nilsson-SKM-HFB}
\begin{array}{llll}
\ket{\frac{5}{2}^+}=&+.504\ket{622}& +.487\ket{642}& +.248\ket{862}\\
                    &+.418\ket{633}& -.383\ket{613}&+\cdots\\[1ex]
\ket{\frac{3}{2}^+}=&+.015\ket{622}& +.642\ket{642}& +.235\ket{862}\\
                    &+.305\ket{611}& -.582\ket{631}&+\cdots \ .
\end{array}
\end{equation}
The SIII-HF calculation gives
\begin{equation}\label{eq:Nilsson-SIII-HF}
\begin{array}{llll}
\ket{\frac{5}{2}^+}=&+.066\ket{622}& +.418\ket{642}& +.180\ket{862}\\
                    &+.755\ket{633}& -.398\ket{613}&+\cdots\\[1ex]
\ket{\frac{3}{2}^+}=&+.134\ket{622}& +.360\ket{642}& +.165\ket{862}\\
                    &+.428\ket{611}& -.642\ket{631}&+\cdots
\end{array}
\end{equation}
and the SIII-HFB with pairing
\begin{equation}\label{eq:Nilsson-SIII-HFB}
\begin{array}{llll}
\ket{\frac{5}{2}^+}=&+.024\ket{622}& +.423\ket{642}& +.159\ket{862}\\
                    &+.775\ket{633}& -.367\ket{613}&+\cdots\\[1ex]
\ket{\frac{3}{2}^+}=&+.156\ket{622}& +.342\ket{642}& +.149\ket{862}\\
                    &+.412\ket{611}& -.636\ket{631}&+\cdots \ .
\end{array}
\end{equation}
In SIII-HF and SIII-HFB the blocked states exhibit more concentration on the
dominant Nilsson orbits $[633]$ and $[631]$. As both have the same nodal
structure in $z$-direction ($N_z=3$) one expects less difference in the
density distribution than in the SkM$^*$ case where both states are superpositions
of several Nilsson orbits with similar amplitudes.
In the SkM$^*$ case the leading component of $\ket{\frac{5}{2}^+}$ is $[622]$
and of $\ket{\frac{3}{2}^+}$ is $[642]$, which implies a change in nodal
structure in $z$-direction. In summary
the polarization of the proton distribution due to the reoccupation of the level
with the unpaired neutron has less effect in the SIII than in the SkM$^*$ case.

In Ref.~\cite{HaFM08} the finite-range microscopic-macroscopic model has been
used to study the problem. The authors find small Coulomb energy differences
similar to our SIII case. Also the decomposition of the last neutron orbits
\footnote{Please note that in Ref.~\cite{HaFM08} the authors seem to
have mixed up the right hand sides of Eqs. (1) and (2).}
shows a mixture of several Nilsson orbits resembling more our SIII states than
the SkM$^*$ states.

We should like to point out that the Nilsson orbit $5/2^+[633]$ that is usually
used to classify the $K^\pi=5/2^+$ ground state band
\cite{BBBB07,RPZM06,BABW03}
does not even contain a single-particle spin $j^\pi=5/2^+$.
Due to positive parity and $\Lambda=|m_l|=3$ the orbital angular momenta
contained in $5/2^+[633]$ are $l=4,6,\dots$.
Thus the lowest possible spin in the $5/2^+[633]$ Nilsson state is $j^\pi=7/2^+$.
This implies that in a core plus valence-neutron picture
the $5/2^+[633]$ state needs to be coupled to the excited $J^\pi=2^+$ state in
$^{228}$Th in order to get the ground state spin $J^\pi=5/2^+$.
In the deformed mean-field description the total angular momentum
$J^\pi=5/2^+$ of the nucleus arises from
both the $5/2^+[633]$ orbital and the underlying
deformed $^{228}$Th core.

In Ref.~\cite{RPZM06,GKAB02} experimental data have been compared with
structure calculations for $^{229}$Th that
use the quasiparticle-phonon model \cite{Solo76} employing a
phenomenological Nilsson mean field and multipole-multipole residual interactions.
A very good description of the  $^{229}$Th level structure has been
achieved by an appropriate fit of the interaction parameters.
It has been found that the coupling of the single-quasiparticle degrees of freedom
to the collective octupole vibrational state of the $^{228}$Th core
is essential to reproduce the parity partner bands observed in experiment.
However, as this model is not self-consistent it can not be used for our
considerations.

In a fully consistent calculation scheme based on a relativistic
density-matrix functional \cite{LitR06} it has been shown that the coupling
to low lying vibrations noticeably improves the description of the
single-particle spectrum around the Fermi surface including the ordering
of the levels. This model is however programmed only for spherical cases.

One has to realize that the situation is not so simple, valence and core nucleons have to be
considered self-consistently and in the next generation of nuclear models,
besides the  projection of the deformed intrinsic state on total spin and particle number,
one also may need to go beyond the mean-field picture by coupling to low-lying
core excitations.

\subsection{Predictive power and accuracy of observables}

In the energy-density functional picture one gives up the explicit knowledge of
a  microscopic Hamilton operator $H(\bs{c})$ acting in many-body Hilbert space.
Its expectation value is replaced by the energy functional
${\cal E}[\,\c,\hat{\rho}\,]$ from which one cannot refer back to the Hamiltonian.
It is important to note that one can also not refer back to a
many-body state $\ket{\Psi}$ that represents an approximation to a true
stationary eigenstate of $H(\bs{c})$. One can of course construct a single
Slater determinant with the operators $a^\dagger_\nu$ which
create the occupied states, but this Slater determinant is more an auxiliary object
that ensures quantum properties like Pauli principle or uncertainty relation.
This Slater determinant misses for example various kinds of typical nuclear
correlations that exist in the true eigenstate.

This raises the important question if one has predictive power for other observables
than the energy. The believe is that observables which can be calculated from
the one-body densities that appear in the energy functionals should be trusted.
In our case we calculate Coulomb and kinetic energies, which are given by
densities that are included in the set of variational variables of the
energy functional  $\E[\,\c,\hat{\rho}\,]$
and therefore should be predicted with high accuracy.

Besides these more general considerations there are also concerns about numerical precision.
In the SkM$^*$ case the Coulomb energy difference
$\Delta V_C$ = $-0.307$~MeV = $(924.854 - 925.161)$~MeV
is a result of subtracting two big numbers that have been calculated numerically.
That means a precision of 10~keV for each of them is desirable.
For SIII-HFB, where $\Delta V_C$=0.001~MeV, a precision of 0.1~keV is needed.
We have checked that we can reach enough numerical precision by sufficient
iteration steps.

But there is also the quest for accuracy.
Let us for example consider the approximate treatment of the exchange term.
Its contribution in HFB-SIII is $-34$~MeV,
a 10\% error means already 3~MeV uncertainty in
$\expect{V_C}=V^{direct}_C+V^{exch}_C$.
On the other hand one would expect that this is mainly a systematic error
that is similar for the two states so that the difference should be less affected,
but 1\% still means 0.3~MeV error.
The actual situation for SIII-HFB is
$\Delta V^{direct}_C=0.29$~keV and $\Delta V^{exch}_C=0.71$~keV
which adds up to the $\Delta V_C$=0.001~MeV listed in Table~\ref{tab:energies}.
In this case the approximate exchange term gives the larger contribution
which weakens strongly the confidence in the calculated value of the amplification.
\begin{table}[htb]
\caption{Violation of the Hellmann--Feynman theorem
$\delta = \alpha_0 \Bigl( \partial E^{tot}(\alpha_0) / \partial \alpha \Bigr)
- E^{Coul}(\alpha_0)$ in HF and
HFB calculations with SIII Skyrme functional for $^{228}$Th.
\label{tab:viol}}
\begin{center}
\tabcolsep=0.8em \renewcommand{\arraystretch}{1.2}%
\begin{tabular}
[c]{c|c|c} \hline\hline
   HF     &  HFB (no cutoff) & HFB (with cutoff)  \\
\hline
 0.02~keV &  0.5~keV         &   200~keV  \\
\hline\hline
\end{tabular}
\end{center}
\end{table}

As discussed in Sec.~\ref{sec:HellmannFeynman} the proper derivation of the equations of motion
from an energy functional is crucial for the validity of the Hellmann--Feynman theorem.
Without this theorem one would calculate numerically the derivative of the energy with respect
to $\alpha$ and create a new source of numerical errors.  We tested the
validity of the Hellmann--Feynman theorem by comparing to numerical derivatives
using a five-point formula. In Table~\ref{tab:viol} the deviations are listed
for the $^{228}$Th ground state of the SIII functional.
In the HF calculation the deviation is within the
numerical uncertainty induced by the five-point formula.
In the HFB calculation one gets also sufficient accuracy
when no cutoff in the quasiparticle subspace contributing to the pairing
interaction is applied.
But the accuracy drops by three orders of magnitude when the density matrices and pairing tensors are
calculated by including contributions from quasiparticle states up to
the cutoff energy of 60\,MeV. The reason is that this truncation does violate the
variational structure of the HFB equations \cite{[Dob09]}. For most
observables, this violation induces small effects, and usually can be
safely neglected, but it does show up in the very demanding
calculation of the Coulomb energy differences.

As discussed before,
another accuracy issue is the fact that our model does not contain projection
on good total spin and sharp particle number.
Also the possibility that configurations could admix that consist of the
$K^\pi=1^-$ band of $^{228}$Th coupled with a single-neutron 5/2$^-$ state cannot be excluded.
These questions have to be the task of future investigations.

\section{Summary}
We have investigated the lowest two states of $^{229}$Th
that are almost degenerate in energy.
Two very successful energy functionals SkM$^*$ and SIII
have been employed in a density-matrix functional theory.  Hartree-Fock and
Hartree-Fock-Bogoliubov calculations have been performed and compared.
The result is that for the SkM$^*$ functional the difference in Coulomb energy
$\Delta V_C$ between excited and ground state ranges from $450$~keV without pairing
to $-300$~keV when pairing effects are included.
On the other hand the SIII-HFB result gives $\Delta V_C=1$~keV only.
The differences in neutron and proton kinetic energies are of similar
size and also quite different for SkM$^*$ and SIII.

Altogether, the nuclear models we used predict amplification factors
$A= \Delta V_C/\omega$,
between the drift of the transition frequency $\delta \omega/\omega$
and the drift $\delta \alpha/\alpha$ in the fine structure constant,
that have absolute values varying between about $10^2$ and $10^4$. We
have pointed out and discussed the fact that the pairing correlations smooth out
polarization effects exerted by the single-particle orbitals.
Therefore, such correlations not only dramatically decrease the anticipated
amplification factors but also make their determination very
uncertain, due to dependence on very detailed properties of the
mean-field and pairing effects.

We have also performed spherical calculations  and conclude that spherical models
should not be consulted as they are too far from reality
to provide serious numbers.

As even the sign of the amplification factor is uncertain,
much more refined calculations are needed that include coupling to low-lying core excitations
and projection on eigenstates with good total angular momentum and particle number.
Before being able to provide reasonably trustable numbers how the transition energy varies
as function of the fine structure constant $\alpha$
one has to make sure that the model reproduces the three low lying rotational
$K^\pi=5/2^+, 3/2^+, 5/2^-$ bands up to $J \approx 9/2$ and the
known electromagnetic transitions within the bands and between them.
This would provide more confidence in the quality of the many-body states
and their Coulomb energy.

In any case the calculations must treat all nucleons (no inert core)
because the whole effect comes from a subtle polarization of the
core protons. Furthermore the model has to be of variational type
in order to make use of the Hellmann--Feynman theorem. Without that
one cannot be sure that the polarization effects caused by the strong
interaction are treated consistently with the necessary accuracy.
Rough estimates and simple minded models are not sufficient.

The experimental endeavor for measuring the drift of the transition frequency
in $^{229}$Th has to be accompanied by substantially improved models
on the nuclear theory side and attempts to gain experimental
information on the Coulomb energies via radii, quadrupole moments
or even form factors.
Without a concerted action of experimental and theoretical efforts
the goal of improved limits on the temporal drift of fundamental
constants cannot be reached.

\begin{acknowledgments}
E. L. acknowledges financial support from the
Frankfurt Institute of Advanced Studies FIAS
and fruitful discussions with E.~E.~Kolomeitsev and D.~N.~Voskresensky.
This work was supported in part by the Polish Ministry of Science
under Contract No.~N~N202~328234 and by the Academy of Finland and
University of Jyv\"askyl\"a within the FIDIPRO programme.
\end{acknowledgments}

%

\end{document}